\documentstyle[12pt,epsfig]{article}
\begin{document}
\vspace{1.0cm}
\begin{flushright}
{hep-ph/9412283

December 1994.}
\end{flushright}
\centerline{ }
\vspace{2.0cm}
\centerline{\bf \Large \bf New Quark Distributions and Semi-Inclusive}
\vspace{0.2cm}
\centerline{\bf \Large \bf Electroproduction on Polarized Nucleons}
\vspace{0.8cm}
\centerline{\large Aram KOTZINIAN
\footnote{Now a visitor at CERN, PPE-Division, CH-1211, Geneva 23,
Switzerland.\\E-mail:ARAM@CERNVM.CERN.CH}}
\vspace{0.8cm}
\centerline{\large Yerevan Physics Institute,}
\vspace{0.2cm}
\centerline{\large Alikhanian Brothers St. 2; AM-375036 Yerevan,}
\vspace{0.2cm}
\centerline{\large Armenia.}
\vspace{2.0cm}

{\bf Abstract.}
\vspace{0.3cm}

{\small
The quark-parton model calculation including the effects of intrinsic
transverse momentum and of all six twist-two distribution functions
of quarks in polarized nucleons is performed.
It is demonstrated that new twist-two quark distribution functions
and polarized quark fragmentation functions can be investigated
in semi-inclusive DIS at leading order in $Q^2$.
The general expression for the cross-section of semi-inclusive DIS
of polarized leptons on polarized nucleons in terms of structure
functions is also discussed.}

\newpage
\section{Introduction}

Deep-inelastic scattering (DIS) of leptons on nucleons provides an
excellent tool for probing the structure of nuclear matter.
The leading-twist momentum and helicity distribution functions (DF)
of quarks in nucleons, $f_1(x)$ and $g_1(x)$, have been intensively
studied.
There exists another independent leading-twist DF, $h_1(x)$ \cite{rs},
\cite{jj}, describing the transverse-spin distribution of
quarks in a transversely-polarized nucleon. These DF's depend on
longitudinal momentum fraction ($x$) carried by quark and are
integrated over its intrinsic transverse momentum ($\vec{k}_T$).
In contrast to chirally even DF's $f_1(x)$ and $g_1(x)$, the
chirally odd DF, $h_1(x)$,
cannot be measured in simple DIS because in this case
transverse spin asymmetries are suppressed at high energies.
It can be measured in the lepton pair production process in
nucleon-nucleon collisions with both nucleons polarized transversely.
To be sensitive to the transversity distribution, it is necessary that
either
the sea quarks are highly polarized or that polarized antinucleons are
used. Thus, at present this experiment seems very difficult.

In semi-inclusive DIS (SIDIS) the transverse-spin DF can be probed if
the transverse polarization of the struck quark
is measured in some way. This ``quark polarimeter" may be provided by
an azimuthal dependence of the fragmentation function (FF) for
transversely-polarized
quarks \cite{col} (Collins effect). Another possibility of
quark polarimetry is based on the observed correlation of flavor and
electric charge of the
fragmenting quark and leading hadron. In Ref. \cite{art}
it was proposed to measure the transverse polarization of quarks
by measuring the polarization of self-analyzing baryons
from fragmentation. An investigation of the
transversity distribution, $h_1(x)$, in SIDIS on transversely-polarized
nucleons has been proposed by the HELP collaboration \cite{help}.

Semi-inclusive DIS on a longitudinally-polarized target has been
considered in Ref. \cite{cm} and \cite{fr}.
It was proposed to measure asymmetries for production of different
types of hadrons to get information about the flavor dependence of
quark helicity DF in nucleons. This kind of measurements was already
performed in the SMC experiment \cite {smc} and is planned
by the HERMES collaboration \cite{herm}.

In theoretical calculations of polarized SIDIS
the intrinsic transverse momentum of the quarks in the nucleon
is usually ignored. For example in \cite{jj1} the polarized
SIDIS cross section integrated over final hadron transverse
momentum ($\vec{P}_T^h$) was considered, keeping higher twist DF and FF.
Because the intrinsic $k_T$ was neglected
and the integration over $\vec{P}_T^h$ was assumed,
the target transverse-spin asymmetry in this case appears only at
twist-three ($\sim 1/Q$). But, as was shown in Ref. \cite{col},
the target transverse-spin asymmetry may exist at twist-two level
in the azimuthal distribution of produced hadron. This asymmetry arises
from the azimuthal dependence of transversely-polarized quark
fragmentation and is sensitive to $h_1(x)$. The parton model picture
of Ref. \cite{col} is not symmetric in the sense that the transverse
momentum of final hadron with respect to the scattered quark was
taken into account but the intrinsic transverse momentum of the
initial quark in nucleon was neglected.

It is known that even in unpolarized SIDIS the effect
of intrinsic momentum can be significant \cite{cahn},\cite{emc}.
For a polarized nucleon the situation is more complicated.
As was shown in Ref. \cite{rs} and \cite{tm} for the nonzero
$k_T$ case, the quark distribution in a polarized nucleon
is described by six DF's already at twist-two (instead of three DF's when
intrinsic $k_T$ effects are neglected).
The three ``new"
DF's relate the transverse (longitudinal) polarization of the quark to
the longitudinal (transverse) polarization of nucleon.
Measurement of these DF's was proposed in the doubly-polarized Drell-Yan
process. Their contribution to polarized DIS appears only at
twist-three ($\sim 1/Q$) \cite{tm1}.

The main subject of this article is the calculation of the polarized
SIDIS cross section in the quark-parton model with
nonzero intrinsic $k_T$.
The lay-out of this paper is as follows: section 2 contains the
derivation and discussion of a general expression for the polarized SIDIS
cross section in terms of structure functions, section 3 contains
a description of the quark-parton model with intrinsic $k_T$
for polarized SIDIS, in the section 4 a final expression for the
SIDIS cross section is derived assuming exponential dependence on
transverse momentum of DF's and FF's, section 5 contains a short
discussion and illustrates possible ways of experimental
investigation of the effects of different DF's and FF's in polarized SIDIS
and finally section 6 contains some concluding remarks.

\section{Structure functions for polarized SIDIS}
\vspace{0.3cm}
Inelastic scattering of polarized leptons on polarized nucleons
$l+N\rightarrow l'+N'+\pi$ was considered a
long time ago by Dombey \cite{domb}. For polarized SIDIS,
Gourdin \cite{go} has counted the number of structure functions and
derived some constraints on them from the requirement of positivity.
To derive a general formula for the cross section these authors used
the decomposition of the hadronic tensor into scalar structure functions
corresponding to different polarizations of the
virtual photon and target.
\footnote{A recent review
on the subject of polarized lepton-nucleon scattering is given by
S. Boffi, C.Giusti and F.D. Pacati \cite{bof}. The author thanks D. von
Harrach for mentioning this article.}

In this section I follow the method used in \cite{domb} and
\cite{go} and present explicit formulae for polarized SIDIS
cross-sections.

The Feynman diagram describing this process
in the one-photon exchange approximation is depicted in Fig. 1.
\begin{figure}[hbt]
\centering
\epsfysize=4cm
\mbox{\epsfbox{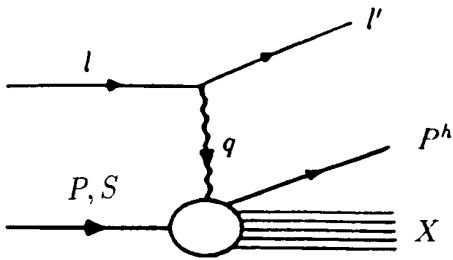}}
\caption{\protect\small Lowest order diagram for SIDIS.}
\label{fig2}
\end{figure}

The standard notation for DIS variables is used:
$l(E)$ and $l'(E')$ are the
momenta (energies) of the initial and final lepton; $\theta_l$ and
$\phi^l_{lab}$ are the scattered electron polar and azimuthal
angles in the laboratory frame; $q=l-l'$ is the exchanged virtual photon
momentum; $\theta_\gamma$ is the virtual photon emission angle; $P$ is the
target nucleon momentum; $P^h$ ($E^h$) is the final hadron momentum
(energy);
$Q^2=-q^2=4EE' /sin^2(\theta_l/2)$; $\nu=P\cdot{q}/M$;
$x=Q^2/2P\cdot{q}$; $y=\nu /E$; $z=P\cdot{P^h_T}/P\cdot{q}$.

For the cross-section one has
\begin{equation}
d^6\sigma^{l+N\rightarrow l'+h+X}
=\frac{1}{4P\cdot{l}}\left(\frac{4\pi\alpha}{Q^2}\right)^{2}
l_{\mu\nu}W^{\mu\nu}(2\pi)^4
\frac{d^{3}l^{'}}{(2\pi)^{3}2E^{'}}
\frac{d^{3}P^{h}}{(2\pi)^{3}2E^{h}}.
\end{equation}
Here the leptonic tensor is given by QED:
\begin{equation}
l_{\mu\nu}=2\left(l_{\mu}l^{'}_{\nu}+l_{\nu}l^{'}_{\nu}
-l\cdot{l^{'}}g_{\mu\nu}+i\lambda
\epsilon_{\mu\nu\alpha\beta}l^{\alpha}q^{\beta}\right),
\end{equation}
where $\lambda$ is the initial lepton helicity ($\mid\lambda\mid\leq 1$).

The hadronic tensor in (1) is defined by
\begin{equation}
W_{\mu\nu}=\sum_X\int\delta^{(4)}(P+q-P^h-P_X)\prod_{i\in{X}}
\frac{d^3P_i}{(2\pi)^{3}2E_{i}}
\langle{P^h},X\mid{J}_{\mu}\mid{P},S\rangle
\langle{P},S\mid{J}_{\nu}\mid{P^h},X\rangle,
\end{equation}
where $S$ is the target nucleon polarization.

The hadronic tensor can be decomposed into a spin-independent
($W_{\mu\nu}^{(0)}$)
and a spin-dependent ($W_{\mu\nu}^{(S)}$) part:
\begin{equation}
W_{\mu\nu}=W_{\mu\nu}^{(0)}+W_{\mu\nu}^{(S)}.
\end{equation}
For a spin-$\frac{1}{2}$ target, $W_{\mu\nu}^{(S)}$ has to be linear in
$S$:
\begin{equation}
W_{\mu\nu}^{(S)}=S^{\rho}W_{\mu\nu\rho}^{(S)}.
\end{equation}
Note that, since $S$ is a pseudovector, $W_{\mu\nu\rho}$ is a
pseudotensor.

It is convenient to consider the hadronic tensor decomposition in terms
of structure functions in the target laboratory frame, with the $z$-axis
chosen in the virtual photon momentum direction and the $x$-axis along
the final hadron transverse momentum $\vec P^{\, h}_T$ (see Fig. 2).
This reference system is referred to
as the laboratory gamma-hadron frame (LGHF).
\begin{figure}[hbt]
\centering
\epsfysize=6cm
\mbox{\epsfbox{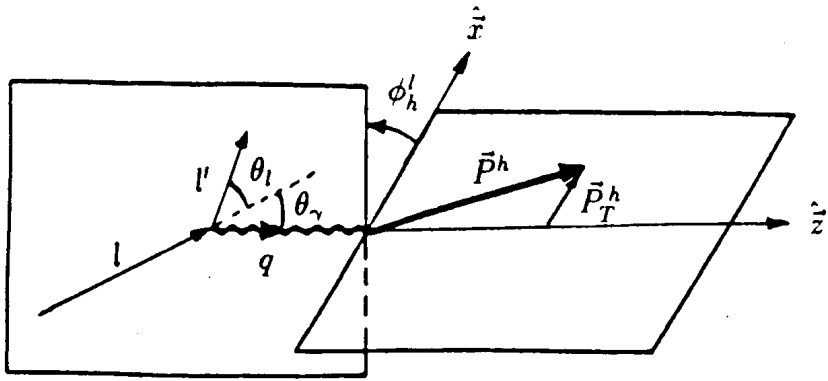}}
\caption{\protect\small Lepton and produced hadron momenta in the LGHF.}
\label{fig1}
\end{figure}

To pass to this reference frame one has to perform three rotations:

1) Rotation around the initial lepton momentum by an angle
$\phi^l_{lab}$, to pass to the lepton scattering plane.

2) Rotation around the normal to the lepton scattering plane by an angle
$-\theta_\gamma$, defining a new $z$-axis coinciding with the virtual
photon momentum direction. This often-used reference frame will be called
the laboratory gamma-lepton frame (LGLF).

3) Rotation around the new $z$-axis by the azimuthal angle of the
produced
hadron ($\phi^h_l$), defining a new $x$-axis coinciding with the produced
hadron transverse momentum direction.

For the following it is important to note that, by definition,
\footnote{Everywhere in this article unit vectors are denoted by a
hat:  $\hat{\vec a}=\vec{a}/|\vec{a}|$.}
$\hat{\vec z}=\hat{\vec q}$, $\hat{\vec x}=\hat{\vec P^h_T}$
are polar vectors and $\hat{\vec y}=[\hat{\vec z}\times\hat{\vec x}]$
is an axial vector.

One of the advantages of the LGHF is that the
hadronic tensor in this frame is independent of the relative azimuthal
angle between the lepton and hadron planes ($\phi^l_h$=$-\phi_l^h$).
To decompose the hadronic tensor into scalar structure functions
the following complete basis for polarization vectors of the
virtual photon ($\epsilon^\mu$) and nucleon ($e^\mu$) is chosen
in the laboratory frame:
\footnote{Of course, this decomposition can be performed in a Lorentz
invariant form.}
\begin{eqnarray}
\epsilon^\mu_0=\frac{1}{Q}(q^3,0,0,q^0),
\;\;\;\;\;\;\;\;\;\;\;\;\;\;\;\
& e^\mu_0 = (1,0,0,0), \nonumber\\
\epsilon^\mu_1=(0,1,0,0),
\;\;\;\;\;\;\;\;\;\;\;\;\;\;\;\;\;\;\;\;\;\
& e^\mu_1 = (0,1,0,0), \nonumber\\
\epsilon^\mu_2=(0,0,1,0),
\;\;\;\;\;\;\;\;\;\;\;\;\;\;\;\;\;\;\;\;\;\
& e^\mu_2 = (0,0,1,0), \\
\epsilon^\mu_3=\frac{q^\mu}{Q},
\;\;\;\;\;\;\;\;\;\;\;\;\;\;\;\;\;\;\;\;\;\;\;\;\;\;\;\;\;\;\;\;\
& e^\mu_3 = (0,0,0,1). \nonumber
\end{eqnarray}

By construction, $\epsilon^\mu_2$ and $e^\mu_2$ are equal to
$\hat{\vec y}$ and thus are axial vectors.
The following closure properties hold for the basis vectors:
\begin{eqnarray}
& & g_{\mu\nu}\epsilon^\mu_a\epsilon^\nu_b=g_{ab}, \nonumber\\
& & g_{\mu\nu}e^\mu_i e^\nu_j=g_{ij}, \nonumber\\
& & g^{ab}\epsilon^\mu_a\epsilon^\nu_b=g^{\mu\nu}, \\
& & g^{ij}e^\mu_i e^\nu_j=g^{\mu\nu}. \nonumber
\end{eqnarray}

Now, expanding both parts of the hadronic tensor over
the complete basis of polarization vectors,
\begin{eqnarray}
&&W_{\mu\nu}^{(0)}=\epsilon^a_\mu\epsilon^b_\nu H^{(0)}_{ab},\nonumber\\
&&W_{\mu\nu\rho}^{(S)}=
\epsilon^a_\mu\epsilon^b_\nu e^i_\rho H^{(S)}_{abi}.\
\end{eqnarray}
one has
\begin{equation}
l_{\mu\nu}W^{\mu\nu}=L^{ab}\left(H^{(0)}_{ab}+S^\rho e^i_\rho
H^{(S)}_{abi}\right),
\end{equation}
where
\begin{equation}
L^{ab}=\epsilon_\mu^a\epsilon_\nu^b l^{\mu\nu}
\end{equation}
can be treated as a virtual-photon polarization density matrix.

Using (7) one can express the scalar structure functions as
\begin{eqnarray}
&&H^{(0)}_{ab}=\epsilon^\mu_a\epsilon^\nu_b W_{\mu\nu}^{(0)},\nonumber\\
&&H^{(S)}_{abi}=\epsilon^\mu_a\epsilon^\nu_b e^\rho_i
W_{\mu\nu\rho}^{(S)}.\
\end{eqnarray}

Let us consider the restrictions imposed by the invariance properties
of $W_{\mu\nu}$ \cite{go} :

- current conservation $\Rightarrow$ structure functions with $a=3$ or
$b=3$ are equal to zero,

- parity conservation $\Rightarrow$ $H_{ab}^{(0)}=0$ if it contains
an odd number of indices 2 and $H_{abi}^{(S)}=0$ if it contains an even
number of indices 2,

- hermiticity of $W_{\mu\nu}$ $\Rightarrow$
$H_{ab}^{(0)}=H_{ba}^{(0)\ast}$
and $H_{abi}^{(S)}=H_{bai}^{(S)\ast}$,

- contribution of $H_{ab0}^{(S)}$ to $W_{\mu\nu}$ is
zero since $e_0^\rho=P^\rho /M$ and $S\cdot P=0$,

- time reversal invariance does not give new constraints on
SIDIS structure functions.

Taking into account these properties one can choose the following
set of independent real structure functions:
five spin-independent
$H_{00}^{(0)}$$, $$H_{11}^{(0)}$, $H_{22}^{(0)}$, $ReH_{01}^{(0)}$,
$ImH_{01}^{(0)}$ and thirteen spin-dependent
$H_{002}^{(S)}$, $ReH_{012}^{(S)}$, $ImH_{012}^{(S)}$,
$eH_{021}^{(S)}$, $ImH_{021}^{(S)}$, $ReH_{023}^{(S)}$,
$ImH_{023}^{(S)}$,
$H_{112}^{(S)}$, $ReH_{121}^{(S)}$, $ImH_{121}^{(S)}$,
$ReH_{123}^{(S)}$, $ImH_{123}^{(S)}$, $H_{222}^{(S)}$.

In the general case these structure functions depend on four variables
$Q^2$ , $\nu$ , $z$ and $P_T^{h2}$.

The lepton momenta in the LGHF have the form:
\begin{eqnarray}
&&l^\mu_{\gamma h}=E\left(1,\sin\theta_\gamma \cos\phi_l^h,
\sin\theta_\gamma \sin\phi_l^h,\cos\theta_\gamma\right),\nonumber\\
&&l^{' \, \mu}_{\gamma h}=E\left((1-y),\sin\theta_\gamma \cos\phi_l^h,
\sin\theta_\gamma \sin\phi_l^h,(1-y)\cos(\theta_\gamma+\theta_l)\right),\
\end{eqnarray}
with
\begin{eqnarray}
&&\cos\theta_\gamma=\sqrt{\frac{Q^2}{Q^2+4M^2x^2}}
\left(1+\frac{2M^2x^2y}{Q^2}\right),\nonumber\\
&&\sin\theta_\gamma=\sqrt{\frac{4M^2x^2}{Q^2+4M^2x^2}
\left(1-y-\frac{M^2x^2y^2}{Q^2}\right)}.\
\end{eqnarray}

By our choice of reference system the dependence on azimuthal angle
$\phi_h^l$ is completely contained in the leptonic tensor.
For ${\cal L}^{ab}$ = $(1-\epsilon)L^{ab}/2Q^2$, one also has
the hermiticity property ${\cal L}^{ab}$ = ${\cal L}^{ba\ast}$ and
from (2), (10) and (12) it follows that
\begin{eqnarray}
& &{\cal L}^{00}=\epsilon ,\nonumber\\
& &{\cal L}^{01}=\sqrt{\frac{\epsilon(1+\epsilon)}{2}}\cos\phi^l_h+
i\lambda\sqrt{\frac{\epsilon(1-\epsilon)}{2}}\sin\phi^l_h ,\nonumber\\
& &{\cal L}^{02}=-\sqrt{\frac{\epsilon(1+\epsilon)}{2}}\sin\phi^l_h+
i\lambda\sqrt{\frac{\epsilon(1-\epsilon)}{2}}\cos\phi^l_h ,\nonumber\\
& &{\cal L}^{11}=\frac{1}{2}+\frac{\epsilon}{2}\cos2\phi^l_h ,\\
& &{\cal L}^{12}=-\frac{\epsilon}{2}\sin2\phi_e^h+
i\frac{\lambda}{2}\sqrt{1-\epsilon^2} ,\nonumber\\
& &{\cal L}^{22}=\frac{1}{2}-\frac{\epsilon}{2}\cos2\phi^l_h ,\nonumber
\end{eqnarray}
where
\begin{equation}
\epsilon=\frac{1}{1+2\frac{\nu^2+Q^2}{Q^2} tg^2 \frac{\theta_l}{2}}
=\frac{2(1-y)-Mxy/E}{1+(1-y)^2+Mxy/E}
\end{equation}
describes the virtual-photon polarization.

One can see that all ``memory" of the scattered lepton azimuthal angle,
$\phi^l_h$, is contained in the polarization density matrix of
virtual photon when SIDIS is described in the LGHF.

Finally, for the product of the hadronic and leptonic tensors
we get
\begin{eqnarray}
&&\frac{1-\epsilon}{2Q^2}l_{\mu\nu}W^{\mu\nu}\nonumber\\
&&=\frac{1}{2}\left(H_{11}^{(0)}+H_{22}^{(0)}\right)
+\epsilon H_{00}^{(0)}+
\sqrt{2\epsilon(1+\epsilon)}ReH^{(0)}_{01}\cos\phi_h^l+
\frac{\epsilon}{2}\left(H_{11}^{(0)}-H_{22}^{(0)}\right)\cos2\phi_h^l
\nonumber\\
&&+\lambda\sqrt{2\epsilon(1-\epsilon)}ImH_{01}^{(0)}\sin\phi_h^l
\nonumber\\
&&-S^1_{(\gamma h)}\left[
\sqrt{2\epsilon(1+\epsilon)}ReH^{(S)}_{021}\sin\phi_h^l+
\epsilon ReH^{(S)}_{121}\sin2\phi_h^l\right]\nonumber\\
&&-S^2_{(\gamma h)}\left[
\frac{1}{2}\left(H_{112}^{(S)}+H_{222}^{(S)}\right)
+\epsilon H_{002}^{(S)}+
\sqrt{2\epsilon(1+\epsilon)}ReH^{(S)}_{012}\cos\phi_h^l+
\frac{\epsilon}{2}\left(H_{112}^{(S)}-H_{222}^{(S)}\right)
\cos2\phi_h^l\right]\nonumber\\
&&-S^3_{(\gamma h)}\left[
\sqrt{2\epsilon(1+\epsilon)}ReH^{(S)}_{023}\sin\phi_h^l+
\epsilon ReH^{(S)}_{123}\sin2\phi_h^l\right]\\
&&+\lambda S^1_{(\gamma h)}\left[-\sqrt{1-\epsilon^2}ImH^{(S)}_{121}+
\sqrt{2\epsilon(1-\epsilon)}ImH^{(S)}_{021}\cos\phi_h^l\right]\nonumber\\
&&-\lambda S^2_{(\gamma h)}\left[
\sqrt{2\epsilon(1-\epsilon)}ImH^{(S)}_{012}\sin\phi_h^l\right]\nonumber\\
&&+\lambda S^3_{(\gamma h)}\left[-\sqrt{1-\epsilon^2}ImH^{(S)}_{123}+
\sqrt{2\epsilon(1-\epsilon)}ImH^{(S)}_{023}\cos\phi_h^l\right],\nonumber
\end{eqnarray}
where $S_{(\gamma h)}^1$, $S_{(\gamma h)}^2$ and $S_{(\gamma h)}^3$
are the target nucleon spin components in the LGHF.
They are related to the target nucleon longitudinal $S^L_{\gamma l}$
and transverse $\vec{S}^T_{\gamma l}=(S^1_{\gamma l},S^2_{\gamma l})$
spin components in the LGLF by
\begin{eqnarray}
&&S_{\gamma h}^1=S_{\gamma l}^T \cos(\phi^S_l-\phi^h_l) ,\nonumber\\
&&S_{\gamma h}^2=S_{\gamma l}^T \sin(\phi^S_l-\phi^h_l) ,\\
&&S_{\gamma h}^3=S_{\gamma l}^L,\nonumber\
\end{eqnarray}
and $\vec {S}_{\gamma l}$ is related to the target polarization
in the laboratory frame by
\begin{eqnarray}
&&S_{\gamma l}^1=S_{lab}^T \cos\theta_\gamma \cos(\phi^S_{lab}
-\phi^l_{lab})+S_{lab}^L \sin\theta_\gamma,\nonumber\\
&&S_{\gamma l}^2=S_{lab}^T \sin(\phi^S_{lab}-\phi^l_{lab}),\\
&&S_{\gamma l}^3=-S_{lab}^T \sin\theta_\gamma
\cos(\phi^S_{lab}-\phi^l_{lab})+S_{lab}^L \cos\theta_\gamma,\nonumber\
\end{eqnarray}
where $S^L_{lab}$ and $\vec{S}^T_{lab}$ are the nucleon spin longitudinal
and transverse components with respect to initial lepton momentum.

Note that in (16)-(18) the dependence of the polarized
SIDIS
cross section on target spin components and produced hadron azimuthal
angle is expressed in explicit form. In principle, it is possible
to separate the different structure functions contribution by using a
``Fourier
analysis" on $\phi^h_l$ for different beam and target polarizations
as was done for the unpolarized case in Ref. \cite{emc}.

Let us now consider the cross section expression in the LGLF
\begin{equation}
\frac{d^6\sigma^{l+N\rightarrow l'+h+X}}
{dxdyd\phi^l_{lab}dzdP_T^{h2}d\phi^h_l}=
\frac{\nu}{4P^h_{\|}}\frac{\alpha^2y}{2Q^4}l_{\mu\nu}W^{\mu\nu},
\end{equation}
and integrate it over $\phi_l^h$. One gets
\begin{eqnarray}
&& \int_0^{2\pi}d\phi_l^h \frac{d^6\sigma^{l+N\rightarrow l'+h+X}}
{dxdyd\phi^l_{lab}dzdP_T^{h2}d\phi^h_l}=
\frac{\nu}{4P^h_{\|}}\frac{2\pi\alpha^2y}{Q^2(1-\epsilon)}
\bigg[\frac{1}{2}\left(H_{11}^{(0)}+H_{22}^{(0)}\right)
+\epsilon H_{00}^{(0)}\nonumber\\
&&+S^2_{(\gamma l)}\sqrt{\frac{\epsilon(1+\epsilon)}{2}}
\left(ReH^{(S)}_{021}-ReH^{(S)}_{012}\right)\\
&&+\lambda S^1_{(\gamma l)}\sqrt{\frac{\epsilon(1-\epsilon)}{2}}
\left(ImH^{(S)}_{021}-ImH^{(S)}_{012}\right)
-\lambda S^3_{(\gamma l)}\sqrt{1-\epsilon^2}ImH^{(S)}_{123}\bigg]
.\nonumber\
\end{eqnarray}

It is interesting to note that, due to the presence of the
third term in the $rhs$ of the last equation, when
integrated over $\phi_l^h$, the cross section can still have a
single target-spin asymmetry.
Furthermore integration over $P_T^{h\, 2}$ and $z$ has to give us
the polarized DIS cross section times the mean hadron multiplicity
($\langle n_h(x,Q^2)\rangle $).

One can show that by integrating over the produced hadron phase space,
the first and last two terms in (22) precisely reproduce the
exact formula for the polarized DIS cross section,
given for example in Ref. \cite{jaf}, and that the following kinematical
sum rules hold:
\begin{eqnarray}
&& \int \frac{d^3P^h}{2E^h}\left[\frac{1}{2}
\left(H^{(0)}_{11}+H^{(0)}_{22}\right)\right]=
\langle n_h(x,Q^2)\rangle  F_1^{DIS}(x,Q^2),\nonumber\\
&& \int \frac{d^3P^h}{2E^h} \left[\frac{1}{2}
\left(H^{(0)}_{11}+H^{(0)}_{22}\right)+H^{(0)}_{00}\right]
\frac{2xQ^2}{Q^2+4M^2x^2}
=\langle n_h(x,Q^2)\rangle  F_2^{DIS}(x,Q^2),\nonumber\\
&& \int \frac{d^3P^h}{2E^h}\left[\frac{2Mx}{Q}
\left(ImH^{(S)}_{021}-ImH^{(S)}_{012}\right)-2ImH^{(S)}_{123}\right]=
\langle n_h(x,Q^2)\rangle  g^{DIS}_1(x,Q^2),\\
&& \int \frac{d^3P^h}{2E^h}\left[\frac{Q}{2Mx}
\left(ImH^{(S)}_{021}-ImH^{(S)}_{012}\right)+2ImH^{(S)}_{123}\right]=
\langle n_h(x,Q^2)\rangle  g^{DIS}_2(x,Q^2),\nonumber\
\end{eqnarray}
where $F_1^{DIS}(x,Q^2)$, $F_2^{DIS}(x,Q^2)$, $g^{DIS}_1(x,Q^2)$
and $g^{DIS}_2(x,Q^2)$ are the structure functions
entering in the spin-independent and spin-dependent part of the polarized
DIS cross section.

As is well known, a single target-spin asymmetry is forbidden in simple
DIS by time reversal invariance.
This means that the following ``sum rule" holds:
\begin{equation}
 \int \frac{d^3P^h}{2E^h}
 \left(ReH^{(S)}_{021}-ReH^{(S)}_{012}\right)=0.
\end{equation}
Thus, one can have a single target-spin asymmetry in
a SIDIS cross section integrated over $\phi_l^h$,
which disappears after integration over $P_T^{h\,2}$ and $z$.
This observation will be confirmed in the simple quark-parton model.

\section{Parton model with intrinsic transverse momentum}
\vspace{0.3cm}
Unpolarized SIDIS is described in a simple way
by the factorized parton model. In this model the lepton knocks out
a quark, which subsequently fragments into hadrons.
Sixteen years ago it was shown by Cahn \cite{cahn}
that nonperturbative effects of the intrinsic transverse momentum ($k_T$)
of the quarks inside the nucleon may induce significant
hadron asymmetries in the relative azimuthal angle $\phi^h_l$.
The EMC experiment \cite{emc} found an azimuthal
asymmetry at the level of up to 15-20\%,
which arises mainly from the effects of the intrinsic $k_T$
of the struck quark with $\langle k_T^2\rangle\geq(0.44GeV)^2$.

Here I generalize the calculation of Ref. \cite{cahn}
to the polarized SIDIS process. Calculations will be
performed in the electron-quark scattering Breit-frame (BF) (Fig. 3),
which can be reached from the LGLF by a Lorentz boost along the $z$-axis.
\begin{figure}[hbt]
\centering
\epsfysize=6cm
\mbox{\epsfbox{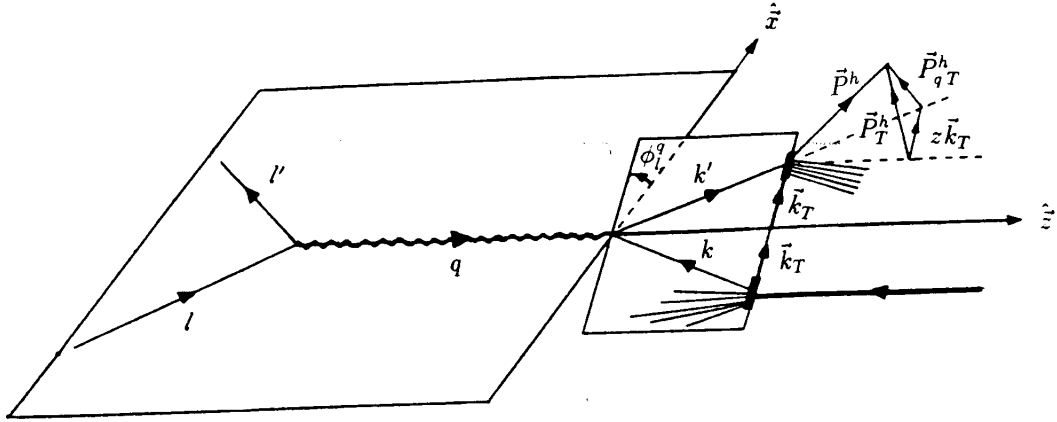}}
\caption{\protect\small The quark-parton model picture for SIDIS
in the BF.}
\label{fig3}
\end{figure}

The azimuthal angle,
$\phi^h_e$, remains unchanged under this transformation,
and, for the virtual photon four-momentum, we have $q^\mu=(0,0,0,Q)$.
In the following calculations the terms $\sim  1/Q^2$ are neglected.
In this approximation
the lepton and quark ($k,k'$) four-momenta in the BF are given by:
\begin{eqnarray}
&&l^\mu_{Br}=\frac{Q}{2}\bigg(\frac{2-y}{y},
\frac{2\sqrt{1-y}}{y}\hat{\vec l}_T,1\bigg),\nonumber\\
&&{l'}^\mu_{Br}=\frac{Q}{2}\bigg(\frac{2-y}{y},
\frac{2\sqrt{1-y}}{y}\hat{\vec l}_T,-1\bigg),\nonumber\\
&&k^\mu_{Br}=\frac{Q}{2}\bigg(1,\frac{2\vec k_T}{Q},-1\bigg),\\
&&{k'}^\mu_{Br}=\frac{Q}{2}\bigg(1,\frac{2\vec k_T}{Q},1\bigg),\nonumber
\end{eqnarray}
where $ \hat{\vec l}_T=(1,0)$ is the unit transverse
two-vector in the lepton transverse momentum direction.
Note that because of a nonzero $\vec k_T$, the electron
and the quark scattering planes do not coincide in general (see Fig. 3,
where the angle between these planes is denoted by $\phi^q_l$).

In the quark-parton model the description of the SIDIS
process is similar to that of double scattering experiments,
in which the polarization of the particle produced after the
first scattering is measured.
To calculate the cross section one has to know:

{\bf 1)} the description of the initial quark state,

{\bf 2)} the noncoplanar polarized $l+q\rightarrow l'+q'$
scattering cross section,

{\bf 3)} the description of the polarized quark fragmentation.

\subsection {\small The initial quark state.}
\vspace{0.3cm}

The polarization of free quarks is most clearly described by the
density matrix of partially polarized fermions, which in the
ultrarelativistic case takes the form \cite{ll}
\begin{equation}
\rho=\frac{1}{2}(\gamma\cdot k)\bigg[1+\gamma^5s_L
+\gamma^5\vec\gamma_T\cdot\vec s_T\bigg],
\end{equation}
where $s_L$ and $\vec s_T$ are the longitudinal and transverse
(with respect to quark three-momentum $\vec k$)
components of twice the quark polarization vector in its rest frame,
$s_L^2+|\vec s_T|^2\leq 1$.

For the initial quark state in the nucleon one has to use, instead
of (2), the following expression:
\begin{equation}
\rho^{q \,\,(in)}_N=\frac{1}{2}{\cal P}^q_N(x,k_T^2)
(\gamma\cdot k)\bigg[1+\gamma^5s_L^{(in)}(x,\vec k_T)
+\gamma^5\vec\gamma_T\cdot\vec s_T^{\,\, (in)}(x,\vec k_T)\bigg],
\end{equation}
where ${\cal P}^q_N(x,k_T^2)$, $s_L^{(in)}(x,\vec k_T)$
and $\vec s_T^{\,\, (in)}(x,\vec k_T)$ are
the probability, longitudinal and transverse polarization
distributions. They can easily be found from the density matrix by
calculating the appropriate trace
\begin{eqnarray}
&&{\cal P}^q_N(x,k_T^2)=\frac{1}{2p^0}tr\left(\rho^{q\,\,(in)}_N
\gamma^0\right),\nonumber\\
&&{\cal P}^q_N(x,k_T^2)s_L^{(in)}(x,\vec k_T)=
\frac{1}{2p^0}tr\left(\rho^{q \,\,(in)}_N\gamma^5\gamma^0\right),\\
&&{\cal P}^q_N(x,k_T^2)\vec s_T^{\,\, (in)}(x,\vec k_T)=
\frac{1}{2p^0}tr\left(\rho^{q\,\, (in)}_N
\gamma^5\vec\gamma_T\gamma^0\right).\nonumber\
\end{eqnarray}

At first sight, it seems that presence of the intrinsic transverse
momentum of the quark can not give a sizable effect on the spin distribution.
Consider, for example, a transversely-polarized quark in the quark-parton
model. In this case the projection
of quark polarization vector onto the nucleon momentum direction is
proportional to $k_T/k_{\|}=2k_T/Q$, and one might conclude that
the contribution of the quark transverse polarization to the
nucleon longitudinal polarization is suppressed at high $Q^2$.
However, the nucleon polarization has to be calculated in its rest
frame, and the factor $1/Q$ will disappear after a Lorentz
transformation to nucleon rest frame.

Consider a simple example for spin transfer from a
polarized nucleon to a quark.
Let $A_\mu$ and $a_\mu$ be the
polarization four-vectors of the nucleon and the quark
in the frame where the nucleon has large momentum (for example in the BF),
and suppose that they are related by
$a_0=\alpha A_0$, $a_3=\alpha A_3$, $a_1=\beta A_1$ and $a_2=\beta A_2$,
with $A_0=P_NS_L/M$, $A_3=E_NS_L/M$, $\vec{A}_T=\vec{S}_T$, where
$\vec{S}$ is the nucleon polarization in its rest frame and $P_N$ ($E_N$)
is the momentum (energy) of the nucleon in the BF.
Now one
can calculate the quark polarization in its rest frame ($\vec{s}$)
assuming that $m_q=xM$. After rotation of the coordinate system of the BF
and a Lorentz boost along the quark momentum one gets:
$s_L=\alpha S_L+\beta\vec{k}_T\cdot\vec{S}_T/m_q$ and
$\vec{s}_T=\beta\vec{S}_T-\alpha\vec{k}_TS_L/m_q$.
Thus, in this ``toy" parton model, the longitudinal spin
of the quark receives two contributions: from both the longitudinal and
transverse spin of the nucleon. It is important to note that neither of these
contributions is suppressed at high $Q^2$. The same behavior is
true for the transverse spin of the quark.

General consideration of the quark DF
in a polarized nucleon in the case of nonvanishing $k_T$ has been
done by Ralston and Soper \cite{rs} and recently by
Tangerman and Mulders \cite{tm}. They have found that at the leading
twist one needs six independent DF's depending on $x$ and $k_T^2$:
$f$,$g_{1L}$, $g_{1T}$, $h_{1T}$, $h_{1L}^\perp$, and $h_{1T}^\perp$ .
The distributions
$s_L^{(in)}(x,\vec k_T)$
and $\vec s_T^{\,\,(in)}(x,\vec k_T)$ are
given by \cite {tm}

\begin{eqnarray}
&&{\cal P}^q_N(x,k_{T}^2) = f(x,k_{T}^2), \nonumber\\
&&{\cal P}^q_N(x,k_{T}^2)\, s_L^{(in)}(x,\vec{k}_{T}) =
g_{1L}(x,k_{T}^2)\,S_L
+  g_{1T}(x,k_T^2)\,\frac{\vec{k}_T\cdot\vec{S}_T}{ m_D}\,,\\
&&{\cal P}^q_N(x,k_{T}^2) \, \vec s_T^{\,\, (in)}(x,\vec{k}_{T}) \,
= h_{1T}(x,k_{T}^2)\,\vec{S}_{T}\, + \left[
h_{1L}^\perp(x,k_{T}^2) S_L +h_{1T}^\perp(x,k_{T}^2)
\frac{\vec{k}_T\cdot\vec{S}_T}{ m_D}\right]\,\frac{\vec{k}_T}
{ m_D},
\nonumber
\end{eqnarray}
where $ m_D$ is an unknown mass parameter, $S_L$ and $\vec S_T$ are
the nucleon
longitudinal and transverse polarization with respect to its momentum.
The ``new" DF's have clear physical interpretation: for example, $g_{1T}$
describes the quark longitudinal polarization in a transversely-polarized
nucleon. It is important to notice that due to this DF
even the initial quark longitudinal spin distribution in a
polarized nucleon exhibits an azimuthal asymmetry.

\subsection{\small Noncoplanar polarized
$l+q\rightarrow l'+q'$ scattering}
\vspace{0.3cm}

Using standard methods of QED \cite{ll}, it is easy to calculate the
cross section of a polarized lepton scattering on a polarized quark
in the one-photon approximation:
\begin{eqnarray}
\frac{d^2\sigma_i^{l+q\rightarrow l'+q'}}{dyd\phi^l_h}&=&
\frac{\alpha^2e_i^2y}{2Q^2}f(x,k_T^2)\bigg[\left(s^2+u^2\right)
\left(1+s_L^{(in)}s'_L\right)
+\left(s^2-u^2\right)\lambda\left(s_L^{(in)}+s'_L\right)\nonumber\\
& &+2su(s_T^{(in)}\cdot s'_T)-4u(s_T^{(in)}\cdot l)(s'_T \cdot l')
-4s(s_T^{(in)}\cdot l')(s'_T \cdot l)\bigg].\
\end{eqnarray}
Here, $e_i$ is the electric charge of the quark in positron charge
units, $s'_L$ is the parameter describing the longitudinal spin
component of the final quark
and $s'_T$ is four-vector describing its transverse spin;
$s$, $t$ and $u$ are the usual Mandelstam variables given by
\begin{eqnarray}
&&s=2MEx\bigg[1-2\sqrt{1-y}\frac{\vec k_T\cdot\hat{\vec l}_T}{Q}
\bigg],\nonumber\\
&&t=-Q^2=-2MExy, \\
&&u=-2MEx(1-y)\bigg[1-\frac{2}{\sqrt{1-y}}
\frac{\vec k_T\cdot\hat{\vec l}_T}{Q}\bigg].\nonumber\
\end{eqnarray}
Note that for noncoplanar $l+q\rightarrow l'+q'$ scattering $s$ and $u$
depend on the relative azimuthal angle between the quark and lepton
scattering planes, $\phi^q_l$, and kinematical corrections of order $1/Q$
will arise due to this dependence.

The transverse polarization four-vectors of the quarks in the BF
in this approximation are
\begin{eqnarray}
&&s^{(in)\, \mu}_T=
(0,\vec s_T^{\,\, (in)},\frac{2}{Q}\vec k_T
\cdot\vec s_T^{\,\, (in)}),\nonumber\\
&&{s'}^\mu_T=
(0,\vec s{\, '}_T,-\frac{2}{Q}\vec k_T\cdot\vec s{\, '}_T).\
\end{eqnarray}

Substituting (29) and (30) into (28) the following expression for
the cross section is obtained:
\begin{equation}
\frac{d^2\sigma_i^{l+q\rightarrow l'+q'}}{dyd\phi^l_h}=
\frac{\alpha^2e_i^2}{2Q^2y}f(x,k_T^2)
\bigg(a+b_Ls'_L+\vec b_T\cdot\vec s{\, '}_T\bigg),
\end{equation}
where
\begin{eqnarray}
&&a=1+(1-y)^2-4(2-y)\frac{\vec k_T\cdot\hat{\vec l}_T}{Q}
+\lambda s_L^{(in)}y\left[2-y-4\sqrt{1-y}
\frac{\vec k_T\cdot\hat{\vec l}_T}{Q}\right],\nonumber\\
&&b_L=\lambda y\left[2-y-4\sqrt{1-y}
\frac{\vec k_T\cdot\hat{\vec l}_T}{Q}\right]
+s_L^{(in)})\left[1+(1-y)^2
-4(2-y)\sqrt{1-y}\frac{\vec k_T\cdot\hat{\vec l}_T}{Q}\right],\nonumber\\
&&\vec b_T=2(1-y)\bigg\{\vec s_T^{\,\, (in)}-
2(\vec s_T^{\,\, (in)}\cdot\hat{\vec l}_T)\hat{\vec l}_T\\
&&\,\,\,\,\,\,\,\,\,\,\,\,+2\frac{2-y}{Q\sqrt{1-y}}\left[
(\vec s_T^{\,\, (in)}\cdot\hat{\vec l}_T)\vec k_T+
(\vec s_T^{\,\, (in)}\cdot\vec k_T)\hat{\vec l}_T-
(\hat{\vec l}_T\cdot\vec k_T)\vec s_T^{\,\, (in)}\right]\bigg\}.
\nonumber\
\end{eqnarray}

The final quark polarization according to the general rules
(see \cite{ll},\S 65) is given by
\begin{eqnarray}
&&s_L^{\, (f)}=\frac{b_L}{a},\nonumber\\
&&\vec s^{\, (f)}_T=\frac{\vec b_T}{a}.\
\end{eqnarray}

{}From (32) and (33) one can see that
the final quark can be transversely polarized
only if the transverse polarization of the initial quark is
not equal to zero. The sideways ($D_{ss}$) and normal ($D_{nn}$)
transverse spin transfer coefficients to leading order in
$1/Q$ are given by
\begin{equation}
D_{nn}=-D_{ss}=\frac{2(1-y)}{1+(1-y)^2+\lambda s_L^{(in)}y(2-y)}.
\end{equation}
The azimuthal angle of the final quark transverse spin ($\phi^{'\,S}_l$)
in this approximation is very simply connected with that of the initial
quark ($\phi^S_l$):
\begin{equation}
\phi^{'\,S}_l=\pi-\phi^S_l
\end{equation}
For unpolarized leptons, expression (34) coincides with the depolarization
factor in Ref. \cite{col}.

Note also that, in contrast with transverse polarization, the
longitudinal polarization of
the final quark is not equal to zero even if the initial quark is
unpolarized but the initial lepton is longitudinally polarized.

For the final quark state before fragmentation, one can now write
the density matrix as
\begin{equation}
\rho^{q \,\,(f)}=\frac{\alpha^2e_i^2}{2Q^2y}f(x,k_T^2)a
\bigg(1+\gamma^5s_L^{\, (f)}+
\gamma^5\vec\gamma_T\cdot\vec s^{\, (f)}_T\bigg).
\end{equation}

One can see that the final quark state has an azimuthal asymmetry in the
relative angle between the lepton and quark scattering planes,
$\phi^q_l$. Part of this asymmetry has a kinematical origin, as
the third term in the expression for $a$ in (32). The azimuthal asymmetry
in unpolarized SIDIS arises from this term \cite{cahn}.
But there also exist terms reflecting the azimuthal anglular dependence of the
initial quark distributions, which are not suppressed at high $Q^2$.

\subsection{\small Polarized quark fragmentation.}
\vspace{0.3cm}

In analogy with the quark probability distribution
in the nucleon one can write the probability of producing
a hadron, $h$, in the polarized quark fragmentation as
\begin{equation}
{\cal P}^h_q(z,\vec P_{q\,T}^h)=\frac{1}{2p^0}tr\left(\rho^{q\,\,(f)}
F(z,\vec P_{q\,T}^h)\gamma^0\right),
\end{equation}
where $F(z,\vec P_{q\,T}^h)$ is the polarized quark fragmentation
function, depending on $z=E^h/E^{q'}$ and the hadron transverse momentum
with respect to the final quark momentum
$\vec P_{q\,T}^h=\vec P^h_T-z\vec k_T$. This FF
can be presented as a sum of spin-independent and spin-dependent parts
\cite{col}.
\begin{equation}
F(z,\vec P_{q\,T}^h)=F^{(0)}(z, P_{q\,T}^{h\,\,2})+
\gamma^5(\vec\gamma_T\cdot[\hat{\vec k'}\times\frac{
\vec P_{q\,T}^h}{ m_F}])F^{(S)}(z, P_{q\,T}^{h\,\,2}),
\end{equation}
where $m_F$ is another unknown mass parameter.
The two terms on the $rhs$ of equation (40) are the only ones
allowed by parity invariance.

Calculating the trace in (37) one gets
\begin{eqnarray}
{\cal P}^h_q(z,\vec P_{q\,T}^h)&=&\frac{\alpha^2e_i^2}{2Q^2y}f(x,k_T^2)
\bigg(aF^{(0)}(z, P_{q\,T}^{h\,\,2})-
(\vec b_T\cdot[\hat{\vec k'}\times\frac{
\vec P_{q\,T}^h}{ m_F}])F^{(S)}(z, P_{q\,T}^{h\,\,2})\bigg)\nonumber\\
&=&
\frac{\alpha^2e_i^2}{2Q^2y}f(x,k_T^2)
\bigg(aF^{(0)}(z, P_{q\,T}^{h\,\,2})+|\vec b_T|\frac{|
\vec P_{q\,T}^h|}{ m_F}F^{(S)}(z, P_{q\,T}^{h\,\,2})\sin\phi_{col}
\bigg),\
\end{eqnarray}
where $\phi_{col}$ is the angle between $\vec P_{q\,T}^h$ and
the final quark transverse polarization.

In contrast to the ordinary FF, $F^{(0)}(z,P_T^2)$,
the spin-dependent part of the polarized quark FF,
$F^{(S)}(z,P_T^2)$,
or, in other words, the analyzing power of polarized fragmentation,
has never been measured.

\section{Results}
\vspace{0.3cm}
In this section the polarized SIDIS cross section
is calculated taking into account all six twist-two
quark DF's and the Collins effect in polarized quark fragmentation.
Kinematical corrections of order $1/Q$ also are kept.

To calculate the cross section one has to integrate
over $\vec{k}_T$ and sum over all quark and antiquark types
the probability of hadron production in final quark fragmentation
(the initial quark probability distribution and
$l+q\rightarrow l'+q'$ scattering cross section is
already included in the final quark density matrix):
\begin{equation}
\frac{d^6\sigma^{l+N\rightarrow l'+h+X}}
{dxdyd\phi^l_{lab}dzdP_T^{h2}d\phi_l^h}=
\sum_q\int d^2k_T{\cal P}^h_q(z,\vec P_{q\,T}^h),
\end{equation}
where $d^2k_T=1/2dk^2_Td\phi^q_l$ and $\phi^q_l$ is the quark azimuthal
angle in the BF.

This integration can be performed analytically if one
supposes that the transverse momentum dependence
in the DF's and FF's may be written in factorized exponential form:
\begin{eqnarray}
&&d_J(x,k_T^2)=\frac{1}{\pi a_J}
exp\left(-\frac{k_T^2}{a_J}\right)d_J(x),\nonumber\\
&&F^{(0,S)}(z,p_T^2)=\frac{1}{\pi a_F^{(0,S)}}
exp\left(-\frac{p_T^2}{a_F^{(0,S)}}\right)F^{(0,S)}(z).\
\end{eqnarray}
Here the index $J=f_1, g_{1L}, g_{1T}, h_{1T},
h_{1L}^{\perp}, h_{1T}^{\perp}$ enumerates
the different DF's. The width of transverse momentum distribution for each
distribution (fragmentation) function
$a_J\sim\langle k_T^2\rangle$
($a_F^{(0,S)}\sim\langle P_{q\,T}^{h\,2}\rangle$)
can in principle depend on $x$ ($z$).

After some calculations the final result looks like:
\begin{equation}
\frac{d^6\sigma^{l+N\rightarrow l'+h+X}}
{dxdyd\phi^l_{lab}dzdP_T^{h2}d\phi^h_l}=
\sum_q\frac{\alpha^2e_q^2}{2Q^2y}\biggl(C_{UP}+C_{DP}+C_{SP}\biggr),
\end{equation}
where the contributions to unpolarized, double (beam and target)
and single (target only) polarized parts of the cross section
are given in the
following three formulae.
\begin{equation}
C_{UP}=A^{(0)}_{f_1}\left[1+(1-y)^2-4y(2-y)
\frac{\alpha_{f_1}^{(0)}P_T^h}{Q}
\cos\phi_l^h \right],
\end{equation}
\begin{eqnarray}
C_{DP}&=&\lambda S^L_{\gamma l} A^{(0)}_{g_{1L}}y\left[2-y-4\sqrt{1-y}
\frac{\alpha^{(0)}_{g_{1L}}P_T^h}{Q}\cos\phi_l^h\right]\nonumber\\
&+&\lambda S^T_{\gamma l}A^{(0)}_{g_{1T}}y\left[\left(2-y-4\sqrt{1-y}
\frac{\alpha^{(0)}_{g_{1T}}P_T^h}{Q}\cos\phi_l^h\right)
\frac{\alpha^{(0)}_{g_{1T}}P_T^h}{ m_D}\cos(\phi_l^h-\phi_l^S)
\right.\\
&&\,\,\,\,\,\,\,\,\,\,\,\,\,\,\,\,\,\,\,\,\,\,\left.
-2\sqrt{1-y}\frac{b^{(0)}_{g_{1T}}}{ m_DQ}\cos(\phi_l^S)\right],
\nonumber\
\end{eqnarray}
\begin{eqnarray}
C_{SP}&=&-2\frac{1-y}{ m_F}\biggl\{S^T_{\gamma l}A^{(S)}_{h_{1T}}
(1-z\alpha_{h_{1T}}^{(S)})P_T^h \sin(\phi_l^h+\phi_l^S)\nonumber\\
&&\,\,\,\,\,\,\,\,\,\,\,\,\,\,\,\,\,\,\,\,\,\,
+S^L_{\gamma l}
\frac{A^{(S)}_{h_{1L}^{\perp}}}{ m_D}\alpha^{(S)}_{h_{1L}^{\perp}}
(1-z\alpha_{h_{1L}^{\perp}}^{(S)})P_T^{h\,2} \sin2\phi_l^h\nonumber\\
&&\,\,\,\,\,\,\,\,\,\,\,\,\,\,\,\,\,\,\,\,\,\,
+S^T_{\gamma l}\frac{A^{(S)}_{h_{1T}^{\perp}}}{ m_D^2}P_T^h
\bigg(\alpha^{(S)\,2}_{h_{1T}^{\perp}}(1-z\alpha_{h_{1T}^{\perp}}^{(S)})
P_T^{h\,2}\cos(\phi_l^h-\phi_l^S)\sin2\phi_l^h
\nonumber\\
&&\,\,\,\,\,\,\,\,\,\,\,\,\,\,\,\,\,\,\,\,\,\,\,\,\,\,\,\,\,\,\,\,\,\,
+(\frac{1}{2}-z\alpha^{(S)}_{h_{1T}^{\perp}})b_{h_{1T}^{\perp}}^{(S)}
\sin(\phi_l^h+\phi_l^S)\bigg)\nonumber\\
&-& 2\frac{2-y}{Q\sqrt{1-y}}\bigg[
S^T_{\gamma l}A^{(S)}_{h_{1T}}\bigg(\alpha^{(S)}_{h_{1T}}
(1-z\alpha^{(S)}_{h_{1T}})P_T^{h\,2}
-zb_{h_{1T}}^{(S)}\bigg) \sin\phi^S_l\\
&&\,\,\,\,\,\,\,\,\,\,\,\,\,\,\,\,\,\,\,\,\,\,\,\,\,
+S^L_{\gamma l}\frac{A^{(S)}_{h_{1L}^{\perp}}}{m_D}
\bigg(\alpha^{(S)\,2}_{h_{1L}^{\perp}}
(1-z\alpha_{h_{1L}^{\perp}}^{(S)})P_T^{h\,2}
+(1-2z\alpha^{(S)}_{h_{1L}^{\perp}})b_{h_{1L}^{\perp}}^{(S)}\bigg)
P_T^h \sin\phi_l^h\nonumber\\
&&\,\,\,\,\,\,\,\,\,\,\,\,\,\,\,\,\,\,\,\,\,\,\,\,\,
 +S^T_{\gamma l}\frac{A^{(S)}_{h_{1T}^{\perp}}}{ m_D^2}
\biggl(\frac{1}{2}z(\alpha^{(S)\,2}_{h_{1T}^{\perp}}P_T^{h\,2}+
2b_{h_{1T}^{\perp}}^{(S)})b_{h_{1T}^{\perp}}^{(S)} \sin\phi_l^S
\nonumber\\
&&\,\,-\alpha^{(S)}_{h_{1T}^{\perp}}[\alpha^{(S)\,2}_{h_{1T}^{\perp}}
(1-z\alpha_{h_{1T}^{\perp}}^{(S)})
P_T^{h\,2}+(2-3z\alpha^{(S)}_{h_{1T}^{\perp}})b_{h_{1T}^{\perp}}^{(S)}]
P_T^{h\,2}\cos(\phi_l^h-\phi_l^S)\sin\phi_l^h
\biggr)
\biggr]
\biggr\}.\nonumber\
\end{eqnarray}

Here the following notation has been adopted
\footnote{To simplify notation the index $q$ has been suppressed
where possible. Obviously, all quantities related with DF (FF)
depend on quark flavor $q$ (and final hadron type $h$).}
\begin{eqnarray}
&&B^{(0,S)}_J=a_F^{(0,S)}+z^2a_J,\,\,\,\,\,\,\,\,
\alpha_J^{(0,S)}=\frac{za_J}{B_J^{(0,S)}},\,\,\,\,\,\,\,
b_J^{(0,S)}=\frac{a_F^{(0,S)}a_J}{B_J^{(0,S)}},\nonumber\\
&&A^{(0,S)}_J=\frac{1}{\pi B_J^{(0,S)}}
exp\left(-\frac{P_T^{h\,2}}{B_J^{(0,S)}}\right)
d_J(x)F^{(0,S)}(z).\
\end{eqnarray}

One can check that both the $\phi^h_l$ and $y$ dependence of the parton
model result (42)-(45) exactly coincide with that required by the
general structure function analysis of section 2.

Integrating (43)-(45) over hadron azimuthal angle one gets
\begin{eqnarray}
\int_0^{2\pi}\,d\phi_l^h\,C_{UP}&=&2\pi
A^{(0)}_{f_1}\bigg(1+(1-y)^2\bigg),\nonumber\\
\int_0^{2\pi}\,d\phi_l^h\,C_{DP}&=&2\pi\lambda y
\left(S^L_{\gamma l} A^{(0)}_{g_{1L}}(2-y)
-2S^T_{\gamma l}\frac{A^{(0)}_{g_{1T}}}{ m_DQ}\sqrt{1-y}
\bigg(\alpha^{(0)\,2}_{g_{1T}}P_T^{h\,2}+b_{g_{1T}}^{(0)}\bigg)\
\cos\phi_l^S
\right),\nonumber\\
\int_0^{2\pi}\,d\phi_l^h\,C_{SP}&=&
4\pi\frac{(2-y)\sqrt{1-y}}{Q m_F}
S^T_{\gamma l}\left\{2A^{(S)}_{h_{1T}}\bigg[\alpha^{(S)}_{h_{1T}}
(1-z\alpha^{(S)}_{h_{1T}})P_T^{h\,2}-zb_{h_{1T}}^{(S)}
\bigg]\right.\\ &&\,\,\left.
+\frac{A^{(S)}_{h_{1T}^{\perp}}}{ m_D^2}
\biggl[z\left(\alpha^{(S)\,2}_{h_{1T}^{\perp}}P_T^{h\,2}+
2b_{h_{1T}^{\perp}}^{(S)}\right)b_{h_{1T}^{\perp}}^{(S)}
\right.\nonumber\\ &&\,\,\left.
-\alpha^{(S)}_{h_{1T}^{\perp}}\bigg(\alpha^{(S)\,2}_{h_{1T}^{\perp}}
(1-z\alpha_{h_{1T}^{\perp}}^{(S)})
P_T^{h\,2}+(2-3z\alpha^{(S)}_{h_{1T}^{\perp}})
b_{h_{1T}^{\perp}}^{(S)}\bigg)P_T^{h\,2}
\bigg]\right\} \sin\phi_l^S.\nonumber\
\end{eqnarray}

As is clear from the last equation, even when
integrated over hadron azimuthal
angle, the SIDIS cross section can still have a single
target spin asymmetry.
Further integration over hadron transverse momentum gives
\begin{eqnarray}
\int d^2P_T^h\,C_{UP}&=&\bigg(1+(1-y)^2\bigg)f_1(x)F^{(0)}(z),\nonumber\\
\int d^2P_T^h\,C_{DP}&=&\lambda y\bigg[S^L_{\gamma l}(2-y)g_{1L}(x)
-2S^T_{\gamma l}\sqrt{1-y}\frac{a_{g_{1T}}}{ m_DQ}g_{1T}(x)
\cos\phi_l^S\bigg]F^{(0)}(z),\\
\int d^2P_T^h\,C_{SP}&=&4\frac{(2-y)\sqrt{1-y}}{Q m_F}
S^T_{\gamma l}\bigg\{h_{1T}(x)
\bigg[\alpha^{(S)}_{h_{1T}}(1-z\alpha^{(S)}_{h_{1T}})
B^{(S)}_{h_{1T}}-zb_{h_{1T}}^{(S)}\bigg]\nonumber\\
&&\,\,+\frac{a_{h_{1T}^{\perp}}}{ m_D^2}h_{1T}^{\perp}(x)
\bigg[\alpha^{(S)}_{h_{1T}^{\perp}}(1-z\alpha^{(S)}_{h_{1T}^{\perp}})
B^{(S)}_{h_{1T}^{\perp}}-zb_{h_{1T}^{\perp}}^{(S)}\bigg]
\bigg\}F^{(S)}(z) \sin\phi_l^S.\nonumber\
\end{eqnarray}

Using definitions (46), one can check that expressions in the
square brackets on the $rhs$ of the last equation are equal to zero.
Thus, in the quark-parton model the single-spin asymmetry disappears
already after the $\vec{P}_T^h$-integration, and sum rule (22) holds.

Finally, let us multiply the two first equations in (48) by $z$,
integrate over $z$ and sum over final hadron types. Using the
well-known momentum sum rule,
\begin{equation}
\sum_h \int dz z F^{(0)}(z)=1,
\end{equation}
the following relations between polarized DIS structure functions
and quark DF's are established:
\begin{eqnarray}
&&F_1^{DIS}(x,Q^2)=\frac{1}{2}\sum_q e^2_q f_1(x),\nonumber\\
&&g_1^{DIS}(x,Q^2)=\frac{1}{2}\sum_q e^2_q g_{1L}(x),\\
&&g_2^{DIS}(x,Q^2)=\frac{1}{2}\sum_q e^2_q
\left(\frac{a_{g_{1T}}}{2M m_Dx}g_{1T}(x)-g_{1L}(x)\right).\nonumber\
\end{eqnarray}

The last equation in (50) shows that both the ``longitudinal",
$g_{1L}(x,k_T^2)$, and ``transverse", $g_{1T}(x,k_T^2)$, parts
of the quark longitudinal-spin distribution give a twist-two contribution
to the $g_2^{DIS}(x,Q^2)$. This result with quark mass and higher-twist
corrections has been found by Tangerman and Mulders
(see eq. 2.29 in Ref. \cite{tm1} and the discussion therein).

\section{Discussion}
\vspace{0.3cm}

In section 4 the cross section for the production of
a single unpolarized hadron was calculated.
However, with a parton model expression (36) for the polarization density
matrix of the final quark, one can also calculate
the cross sections for the production of polarized self-analyzing
baryons or multi-hadron states (double-particle Collins effect,
handedness, ...).

To investigate transverse spin distributions, it seems more convenient
to use an unpolarized lepton beam and a polarized target because, in this
case, six of the structure functions in (18) do not contribute to the
cross section, i.e., $C_{DP}=0$ in the parton model.

I would like to make several remarks here.

In the parton model calculations of section 4,
terms $\sim 1/Q$ have been kept. In general
to get self-consistent results to this order in $1/Q$ one has to
take into account also first-order QCD radiative corrections and
twist-three DF's and FF's. However, as one can conclude from the EMC
data analysis \cite{emc}, the most important contribution to azimuthal
dependence comes from kinematical effect of the intrinsic
transverse momentum (eq. 43), as calculated by Cahn \cite{cahn}.

In some articles concerning polarized DIS and SIDIS
only $f_1(x)$, $g_1(x)$ and $h_1(x)$ are considered.
They are related to the DF's used in section 3 by $\vec{k}_T$-
integration
\begin{eqnarray}
&&f_1(x)=\int d^2k_Tf_1(x,k_{T}^2),\nonumber\\
&&g_1(x)=\int d^2k_Tg_{1L}(x,k_{T}^2)=g_{1L}(x),\\
&&h_1(x)=\int d^2k_T\left[h_{1T}(x,k_{T}^2)+
\frac{k_T^2}{2 m_D^2}h_{1T}^\perp(x,k_{T}^2)\right]=h_{1T}(x)+
\frac{a_{h_{1T}^{\perp}}}{2 m_D^2}h_{1T}^{\perp}(x).\nonumber\
\end{eqnarray}
The DF's $g_{1T}(x)$ and $h_{1L}(x)$ do not give
contributions to $g_1(x)$ and $h_1(x)$.

However, the nonperturbative effects of intrinsic transverse momentum
play an important role for polarized SIDIS when
dependence on azimuthal angle of the produced hadron is considered.
First of all, even in the zeroth approximation in $1/Q$, all six
twist-two DF's contribute to the cross section in this case
and the dependence on the produced hadron azimuthal
angle, though explicitly calculable, is rather complicated.

The quark-parton model calculations of Ref. \cite{col},
\cite{cm} and \cite{fr} has been performed, neglecting the effects of quark
intrinsic transverse momentum. This corresponds to a $\delta(\vec{k})$
distribution for intrinsic $k_T$ or, in our notation, to
$a_J=0$ and so $\alpha_J^{(0,S)}=0$ and $b_J^{(0,S)}=0$.
This rather arbitrary assumption is not true in the case of
$f_1(x,k_T^2)$. Experiments indicate that $a_{f_1}$ can reach a
value of $\approx0.4(GeV/c)^2$ (see \cite{emc} and \cite{stir}).

Let us consider (44)-(47) in the zeroth approximation in $1/Q$.
In this approximation $\phi_{col}=
\phi^h_l+\phi^S_l-\pi$. In the original paper by Collins \cite{col}
$a_J=0$ was assumed and only the first term on the $rhs$ of eq. (45)
was obtained. But, as one can see from (42)-(45), the azimuthal
dependence of the polarized SIDIS cross section
exhibits more complicated behavior already at zeroth order in $1/Q$
when intrinsic $k_T$ effects are taken into account.

I would like to stress here that, in contrast to parton model result for the
azimuthal asymmetry in unpolarized SIDIS \cite{cahn},
in the case of polarized SIDIS, the azimuthal dependence
appears at leading order in $Q^2$. Thus, in polarized SIDIS
there are not only $\sim 1/Q$ effects of kinematical origin
(as in the unpolarized case) but also leading ones coming from
the azimuthal dependence of the initial quark distribution and/or polarized
fragmentation. Suppose for example that the polarized part of the FF is equal
to zero. Then $C_{SP}=0$, but one can still have a leading-order
azimuthal asymmetry in the second term of the $rhs$ of
(44) due to the nonzero longitudinal spin of the initial quark in
the transversely-polarized nucleon (if $g_{1T}\neq 0$).

\subsection {\small Examples of the target spin asymmetries.}
\vspace{0.3cm}

Suppose that the SIDIS cross section is measured for
different beam helicities and target polarizations.
Then, using the specific forms of the $\phi^h_l$, $\phi^S_l$, $P_T^2$
and $y$ dependence of the different terms in (42)-(45) one can separate
the contribution of the different DF's and FF's.

For example, consider the case of a longitudinally-polarized
(in the laboratory frame)
lepton and target. The target longitudinal polarization
asymmetry with fixed lepton beam helicity, defined as
\footnote{For simplicity 100\% target polarization is considered.}
\begin{equation}
{\cal A}_{L}=\frac{d\sigma^{\rightarrow}-d\sigma^{\leftarrow}}
{d\sigma^{\rightarrow}+d\sigma^{\leftarrow}},
\end{equation}
to zeroth order in $1/Q$ is given by
\begin{equation}
{\cal A}_{L}=
\frac{{\cal C}_{g_{1L}}^{(0)}+{\cal C}_{h_{1L}^{\perp}}^{(S)}}
{{\cal C}_{f_1}^{(0)}},
\end{equation}
where the contribution of different DF's and FF's are:
\begin{eqnarray}
&&{\cal C}_{f_1}^{(0)}=
\bigg[1+(1-y)^2\bigg]\sum_q e^2_q A^{(0)}_{f_1}, \nonumber\\
&&{\cal C}_{g_{1L}}^{(0)}=
\lambda y(2-y) \sum_q e^2_q A_{g_{1L}}^{(0)},\\
&&{\cal C}_{h_{1L}^{\perp}}^{(S)}=-2(1-y) \sin2\phi_l^h
\frac{P_T^{h\,2}}{m_Dm_F}\sum_q e_q^2 \alpha^{(S)}_{h_{1L}^{\perp}}
(1-\alpha^{(S)}_{h_{1L}^{\perp}})A^{(S)}_{h_{1L}^{\perp}}.\nonumber\
\end{eqnarray}
As is clear from(54), one can separate the contributions of
${\cal C}_{g_{1L}}^{(0)}$ and ${\cal C}_{h_{1L}^{\perp}}^{(S)}$ by
measuring the target longitudinal-spin asymmetry for different values of
$\phi_l^h$, $P_T^h$ and $y$. If the experiment shows that
${\cal C}_{h_{1L}^{\perp}}^{(S)}\neq 0$, then one can conclude
that both the twist-two DF,
$h_{1L}^{\perp}$, and FF, $F^{(S)}$, are nonzero.
Thus, in principle, it is possible to investigate the Collins effect
(a spin dependent FF) with a longitudinally polarized beam and target
by analyzing existing SMC semi-inclusive data in different bins
of $\phi^h_l$, as was done without such binning in \cite{smc}.

Analogously one can consider the target transverse polarization
asymmetry with a fixed helicity of the lepton beam, defined by
\begin{equation}
{\cal A}_{T}=\frac{d\sigma^{\uparrow}-d\sigma^{\downarrow}}
{d\sigma^{\uparrow}+d\sigma^{\downarrow}}.
\end{equation}
In the zeroth order on $1/Q$
\begin{equation}
{\cal A}_{T}=
\frac{{\cal C}_{g_{1T}}^{(0)}+{\cal C}_{h_{1T}}^{(S)}
+{\cal C}_{h_{1T}^{\perp}}^{(S)}}{{\cal C}_{f_1}^{(0)}},
\end{equation}
with
\begin{eqnarray}
&&{\cal C}_{g_{1T}}^{(0)}=
\lambda y(2-y)\frac{P^h_T}{m_D}\cos(\phi_l^h-\phi_l^S)
\sum_q e^2_q \alpha_{g_{1T}}^{(0)}A_{g_{1T}}^{(0)},\nonumber\\
&&{\cal C}_{h_{1T}}^{(S)}=-2(1-y)\frac{P_T^h}{m_F}\sin(\phi_l^h+\phi_l^S)
\sum_q e^2_q (1-z\alpha_{h_{1T}}^{(S)})A_{h_{1T}}^{(S)},\\
&&{\cal C}_{h_{1T}^{\perp}}^{(S)}=-2(1-y)\frac{P_T^h}{m_Fm_D^2}
\sum_q e^2_q A^{(S)}_{h_{1T}^{\perp}}
\bigg[
(\frac{1}{2}-z\alpha^{(S)}_{h_{1T}^{\perp}})b_{h_{1T}^{\perp}}^{(S)}
\sin(\phi_l^h+\phi_l^S) \nonumber\\
&&\,\,\,\,\,\,\,\,\,\,\,\,\,\,\,\,\,\,\,\,\,\,\,\,\,\,\,\,\,\,\,\,\,\,\,
+\alpha^{(S)\,2}_{h_{1T}^{\perp}}(1-z\alpha_{h_{1T}^{\perp}}^{(S)})
P_T^{h\,2}\cos(\phi_l^h-\phi_l^S)\sin2\phi_l^h \bigg].\nonumber\
\end{eqnarray}
The contributions of
${\cal C}_{h_{1T}}^{(S)}$ and ${\cal C}_{h_{1T}^{\perp}}^{(S)}$
arising from spin-dependent part of FF can be suppressed if
${\cal A}_T$ is measured in small bins around
$\phi_l^h\approx\phi^S_l\approx 0$.
In these bins the asymmetry is given by
\begin{equation}
{\cal A}_{T}\approx\frac{\lambda y(2-y)P^h_T
\sum_q e^2_q \alpha_{g_{1T}}^{(0)}A_{g_{1T}}^{(0)}}{[1+(1-y)^2]
m_D\sum_q e^2_q A_{f_1}^{(0)}}.
\end{equation}
In this way it seems feasible to investigate the twist-two DF, $g_{1T}$,
already with existing SMC semi-inclusive data \cite{smc1}
\footnote{In the Ref. \cite{smc1} only inclusive data was analyzed.}
on a transversely polarized nucleon.

As it is clear from the preceding examples, with a sufficient amount of
experimental data it is possible to separate the contributions of the
different DF's and FF's by considering appropriate asymmetries in
different
bins of $\phi^h_l$ and $\phi^S_l$ or using the Fourier-analysis method.
It is also possible to perform a flavor analysis, as proposed
in Ref. \cite{help}-\cite{fr}, by measuring SIDIS asymmetries on proton
and deuteron targets for different types of final hadron.

Unfortunately, there are no measurements or theoretical calculations
of the ``new" DF's $g_{1T}$, $h_{1T}^\perp$ and $h_{1L}^\perp$
or the polarized part of the FF's, and numerical estimation of
different asymmetries in polarized SIDIS is now impossible.

\section{Conclusions}
\vspace{0.3cm}
In this paper polarized SIDIS has been considered
in the quark-parton model with a nonzero intrinsic $k_T$,
taking into account all six leading-twist DF's and the Collins effect in
the FF's and keeping kinematical corrections of order $\sim 1/Q$.
It is shown that already at zeroth order in $1/Q$ the effects of
intrinsic $k_T$ are not negligible: all six twist-two DF's with the
polarized part of the FF's give contributions to the azimuthal dependence
of the cross section. A study of appropriate asymmetries in polarized
SIDIS will allow an investigation of the different DF's and FF's
at twist-two. It is
possible to start this type of analysis already with existing SMC
semi-inclusive data.

It was also noticed that a single target-spin asymmetry
(with an unpolarized
lepton beam) can exist in the SIDIS cross section when
integrated over final hadron azimuthal angle. In the quark-parton
model this asymmetry is of order $1/Q$, and disappears after $P_T^{h\,2}$
integration.

New polarized SIDIS experiments like thoss proposed by
the HELP and HERMES collaborations will certainly be very instructive
for the investigation of the new DF's and FF's.

\section*{Acknowledgements}
\vspace{0.3cm}
I wish to thank G. Altarelli, X. Artru, E. Berger, L.Dick,
A. Efremov, P.G. Ratcliffe, G. Veneziano and B. Vuaridel
for useful discussions.

This work was supported in part by Professor Louis Dick and Geneva
University.

\end{document}